\documentclass[pre,onecolumn]{revtex4}
\usepackage{amssymb}
\usepackage{amsmath}
\usepackage{graphicx}
\begin{document}
\title{Repetitive Infection Spreading and Directed Evolution in the Susceptible-Infected-Recovered-Susceptible Model}
\author{Hidetsugu Sakaguchi and Keito Yamasaki}
\address{Interdisciplinary Graduate School of Engineering Sciences, Kyushu
University, Kasuga, Fukuoka 816-8580, Japan}
\begin{abstract}
We study two simple mathematical models of the epidemic. At first, we study the repetitive infection spreading in a simplified SIRS model including the effect of the decay of the acquired immune. The model is an intermediate model of the SIRS model including the recruitment and death terms and the SIR model in which the recovered population is assumed to be never infected again. When the decay rate $\delta$ of the immune is sufficiently small, the multiple infection spreading occurs in spikes. The model equation can be reduced to be a map when the decay rate $\delta$ is sufficiently small, and the spike-like multiple infection spreading is reproduced in the mapping. The period-doubling bifurcation and chaos are found in the simplified SIRS model with seasonal variation. The nonlinear phenomena are reproduced by the map. Next, we study coupled SIRS equations for the directed evolution where the mutation is expressed with a diffusion-type term. A kind of reaction-diffusion equation is derived by the continuum approximation for the infected population $I$. The reaction-diffusion equation with the linear dependence of infection rate on the type space has an exact Gaussian solution with a time-dependent average and variance. The propagation of the Gaussian pulse corresponds to the successive transitions of the dominant variant. 
\end{abstract}
\maketitle
\section{Introduction}
  The nonlinear dynamics of epidemic models has been studied by many authors, since Kermack and McKendrick proposed a mathematical model for epidemics in 1927~\cite{KM}. The model describes the population dynamics of $S$ (susceptible), $I$ (infected), and $R$ (recovered), and therefore, it is called the SIR model. 
If the population $E$ of the exposed population before the appearance of symptoms is included, the SIR model is generalized to the SEIR model~\cite{Anderson}. In the SIR model, the total population is conserved. In a lengthy period, the total population can grow or decay. The endemic SIR model includes the effect of the growth and decay~\cite{Hethcote}. If the transition term from the recovered population to the susceptible population is added to the endemic SIR model, the SIRS model (susceptible-infected-recovered-susceptible model) is derived~\cite{Hethcote}. Furthermore, various modified SIR models such as the SIS model (susceptible-infected-susceptible model), SIRD model (susceptible-infectious-recovered-deceased model), SEIS (susceptible-exposed-infected-susceptible model), and so on have been studied for the analysis of the spread of disease~\cite{Murray, Diekmann, Capasso}. Nobel performed a numerical simulation of the geographic and temporal development of plagues using the mathematical model with diffusion terms~\cite{Nobel}. The infection spreading on various lattices and networks has been a topic of statistical mechanics and complex network science~\cite{Katori, Pastor, Odor}. We studied the slow decay of infection in the SIR model with random infection rate on lattices~\cite{Sakaguchi}. 
 The epidemic models were also used to study vaccination strategy~\cite{Shulgin, Laguzet}. The recent COVID-19 pandemic has excited the study of mathematical models of epidemic~\cite{Science}. The behaviors in the post-pandemic period are predicted using various epidemic models~\cite{Kissler, Are}.

The SIR model is expressed as~\cite{KM,Murray}.  
\begin{eqnarray}
\frac{dS}{dt}&=&-\beta SI,\nonumber\\
\frac{dI}{dt}&=&\beta SI-\gamma I,\nonumber\\
\frac{dR}{dt}&=&\gamma I, \label{SIR}
\end{eqnarray}
where $S$, $I$, and $R$ are respectively susceptible, infected, and recovered populations. The parameters $\beta$ and $\gamma$ denote infection and recovering rates. The total population $S+I+R$ is conserved. 
 In this model, the infection spreading occurs only once, and the infected population decays monotonously after the infection spreads. Multiple spreading often occurs in the pandemic such as COVID-19. 
There are several reasons for the multiple infection spreading. Regulations such as the lockdown and the relaxation might be related to the multiple spreading. The generation of new variants is another reason for the waves~\cite{Dutta}. Here, we consider the effect of the decay of the acquired immune. 

The SIRS model including the effect of the day of the acquire immune is expressed as  
\begin{eqnarray}
\frac{dS}{dt}&=&b-\mu S-\beta SI+\delta R,\nonumber\\
\frac{dI}{dt}&=&\beta SI-\gamma I,\nonumber\\
\frac{dR}{dt}&=&\gamma I-\delta R, \label{SIRS}
\end{eqnarray}
where $b$ is the recruitment rate of the susceptible population, $\mu$ is the death rate, and $\delta$ is the rate at which the recovered population becomes susceptible because the effect of the acquired immune wears off. In this model, $S+I+R$ is not conserved. 

In this paper, we study a simplified SIRS model with $b=\mu=0$. The simplified SIRS model is an intermediate model of the SIR model and the SIRS model.  
The simplified SIRS model exhibits nontrivial nonlinear dynamics such as the spike-like infection waves. In Sec.~2, we study the repetitive waves of infection spreading. The simplified SIRS model is reduced to the SIR model when the transition rate $\delta$ from the recovery population to the susceptible population is zero. Although the infection spreading occurs only once in the SIR model, the repetitive infection spreading occurs, and the peak amplitude of $I(t)$ decays in time in the simplified SIRS model even if the parameter $\delta$ is sufficiently small. The interval between the successive spikes increases and the decay rate of peak amplitude of $I(t)$ decreases when $\delta$ approaches 0. We will show that the nonlinear dynamics of the differential equation can be approximated by a discrete map. In Sec.~3, we consider the simplified SIRS model in case the infection rate changes periodically. The repetitive infection waves is maintained owing to the periodic change of the infection rate, and the periodic and chaotic dynamics appear. The nonlinear dynamics can also be approximated by the discrete map. In Sec.~4, we study a one-dimensional lattice model that takes the form of coupled equations of the SIRS model. We find the directed evolution to higher infectivity and lower virulence occurs in this simple model. We show that the partial differential equation obtained by the continuum approximation has a Gaussian solution with a time-dependent average and variance. In Sec.~5, we summarize the results. 
\section{Repetitive infection spreading in a simplified susceptible-infected-recovered-susceptible model}
Our simplified SIRS model without the recruitment and death terms is expressed as
\begin{eqnarray}
\frac{dS}{dt}&=&-\beta SI+\delta R,\nonumber\\
\frac{dI}{dt}&=&\beta SI-\gamma I,\nonumber\\
\frac{dR}{dt}&=&\gamma I-\delta R, \label{MSIR}
\end{eqnarray}
This model is an intermediate model of Eq.~(\ref{SIR}) and Eq.~(\ref{SIRS}).
The total population $S+I+R$ is conserved similarly to Eq.~(\ref{SIR}). Since we study the case of sufficiently small $\delta$, our model is close to the SIR model. However, the small perturbation terms $\delta R$ change the system behavior drastically. That is, only one infection wave appears in Eq.~(\ref{SIR}), but repetitive infection waves appear in the model Eq.~(\ref{MSIR}). We will show the mechanism of the repetitive infection waves using an approximate map derived from Eq.~(\ref{MSIR}).    
 
Since the total population is constant $S+I+R=N$, Eq.~(\ref{MSIR}) is reduced to\begin{eqnarray}
\frac{dS}{dt}&=&-\beta SI+\delta (N-S-I),\nonumber\\
\frac{dI}{dt}&=&\beta SI-\gamma I. \label{MSIR2}
\end{eqnarray}
There are two stationary solutions: $S=N,\;I=0$ and $S=\gamma/\beta,\;I=\delta(\beta N-\gamma)/\{\beta(\delta+\gamma)\}$. If $\beta N>\gamma$, the first solution of no infection becomes unstable and the second solution becomes a stable solution. The stability of the second solution is determined by eigenvalues of the Jacobian matrix:
\[A=\left (\begin{array}{cc}
-\delta(\beta N-\gamma)/(\delta+\gamma)-\delta, & -\gamma-\delta\\
 \delta(\beta N-\gamma)/(\delta+\gamma), & 0\end{array}\right ).\]
The trace $t$ of the matrix is $t=-\delta(\beta N-\gamma)/(\delta+\gamma)-\delta$ and the determinant $D$  is $D=\delta(\beta N-\gamma)$.
Because $\beta N>\gamma$, $D>0$, and $t<0$, the real part of the eigenvalues is negative, and the solution is stable. If $D>t^2/4$, the eigenvalues are complex. If $\delta$ is sufficiently small, $D>t^2/4$ is satisfied and the damping oscillation occurs. The critical value of $\delta_c$ between the damping oscillation and exponential decay is 
\[\delta_c=\frac{4(\beta N-\gamma)(\delta+\gamma)^2}{(\beta B-\gamma+\delta +\gamma)^2}.\]
\begin{figure}[h]
\begin{center}
\includegraphics[height=4cm]{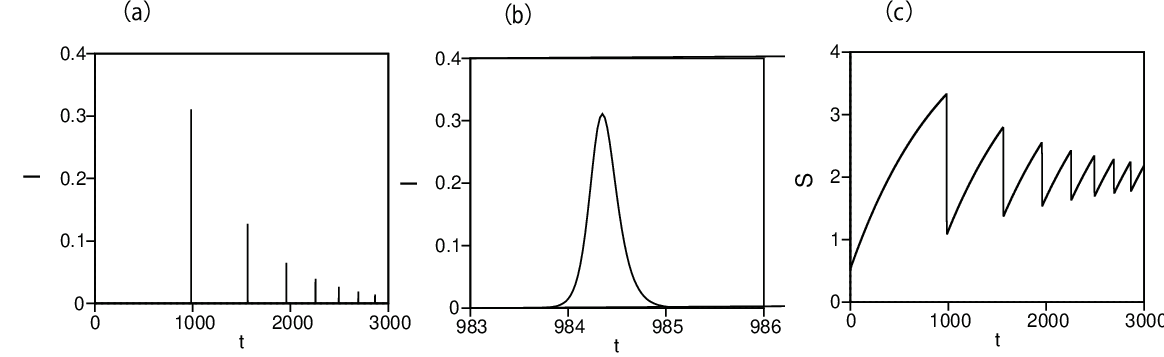}
\end{center}
\caption{(a) Time evolutions of $I(t)$ at $\delta=0.001$, $N=5$, $\beta=10$, and $\gamma=20$. The initial condition is $S(0)=4.999$ and $I(0)=0.001$. (b) Magnification of Fig.~2(a) for $983<t<986$. (c) Time evolution of $S(t)$.}
\end{figure}

For sufficiently small $\delta$, the damping oscillation of $I(t)$ seems to be a spiking oscillation. Figure 1(a) shows the time evolutions of $I(t)$ at $\delta=0.001$, $N=5$, $\beta=10$, and $\gamma=20$. The initial condition is $S(0)=4.999$ and $I(0)=0.001$. The spike-like infection spreading repeats in this long timescale. Figure 1(b) is the magnification of Fig.~1(a) for $983<t<986$. The spreading and ending of infection occur continuously in this short timescale. Figure 1(c) shows the time evolution of $S(t)$ in the long timescale. After the infection spreads, $S(t)$ decreases rapidly, however,  $S(t)$ increases slowly by the term of $\delta R$ after the end of infection spreading. 
When $\beta S(t)$ is larger than $\gamma$, the infection spreading starts again. 
\begin{figure}[h]
\begin{center}
\includegraphics[height=4cm]{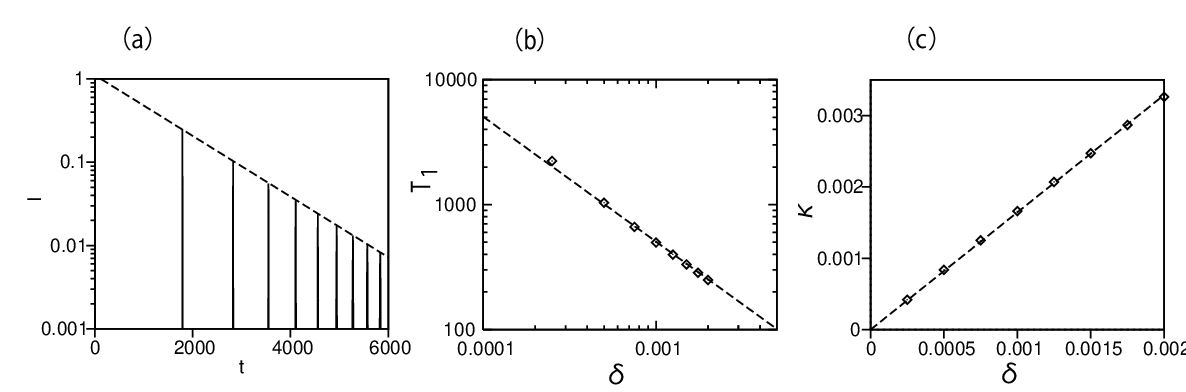}
\end{center}
\caption{(a) Time evolutions of $I(t)$ at $\delta=0.0005$, $N=5$, $\beta=2$, and $\gamma=4$. The initial condition is $S(0)=4.999$ and $I(0)=0.001$. (b) Time interval between the second and third peaks of $I(t)$ for several $\delta$'s in the double-logarithmic scale. The dashed line is $0.505/\delta$. (c) Decay constant $\kappa$ obtained by the fitting of the first seven peaks of $I(t)$ for several $\delta$'s. The dashed line is $1.644\delta$.}
\end{figure}

Figure 2(a) shows the time evolution of $I(t)$ at $N=5$, $\beta=2$, $\gamma=4$, and $\delta=0.0005$ in a semi-logarithmic scale. The time intervals between the two successive peaks of $I(t)$ decrease with $t$. The dashed line is $1.11e^{-\kappa t}$ with $\kappa=0.000839$. That is, the peak amplitude of $I(t)$ decays exponentially initially, although $I(t)$ approaches a constant value at $t=\infty$.  Figure 2(b) shows the time interval between the second and third peaks of $I(t)$ for several $\delta$'s in the double-logarithmic scale. (The first peak appears at $t\simeq 0$.) The dashed line is $0.505/\delta$. The time intervals increase as $\delta$ approaches zero. Figure 2(c) shows the decay constant $\kappa$ obtained by the fitting of the first seven peaks of $I(t)$ for several $\delta$'s. The dashed line is $1.644\delta$. As $\delta$ approaches 0, the decay constant decreases linearly with $\delta$.

We can study the repetitive infection spreading with a map if $\delta$ is sufficiently small.  In the original SIR model Eq.~(\ref{SIR}), a quantity $Q=S+I-(\gamma/\beta)\log S$ takes a constant value in the time evolution. After the end of infection, $I(\infty)=0$ and $S(\infty)$ satisfies 
\begin{equation}
S(\infty)-(\gamma/\beta)\log S(\infty)=S(0)-(\gamma/\beta)\log S(0), \label{SF}
\end{equation}
if $I(0)$ is sufficiently small. From this relation, the final susceptible population $S(\infty)$ is numerically evaluated from the initial population $S(0)$. Similarly, it can be shown that $I$ takes a peak value 
\[S(0)-(\gamma/\beta) \log S(0)-\gamma/\beta+(\gamma/\beta)\log(\gamma/\beta)\]
when $\partial I/\partial S=\partial \{Q-S+(\gamma/\beta)\log S\}/\partial S=0$ or $S=\gamma/\beta$ by the conservation law of $Q=S+I-(\gamma/\beta)\log S$.  

Next, we consider the dynamics in the long timescale. We assume that the $n$th infection spreading starts at $t_{n0}$ and ends at $t_{n\infty}$. 
The susceptible population $S(t_{n\infty})$ just after the $n$th infection spreading is determined by the susceptible population $S(t_{n0})$ just before the $n$th infection spreading by Eq.~(\ref{SF}), where $S(t_{n0})$ and $S(t_{n\infty})$ correspond respectively to $S(0)$ and $S(\infty)$ in Eq.~(\ref{SF}). 
If $S(t_{n\infty})$ is determined by Eq.~(\ref{SF}) from $S(t_{n0})$, $I(t_{n\infty})=I(t_{n0})$ is satisfied since $Q=S+I-(\gamma/\beta)\log S$ takes the same value at $t_{n0}$ and $t_{n\infty}$.  
Hereafter, we assume approximately $t_{n0}=t_{n\infty}=t_n$, since $t_{n\infty}-t_{n0}$ is considered to be negligible in the long-time dynamics as shown in Fig.~2(a).    
After the end time $t_{n\infty}$ of the $n$th infection spreading, $S(t)$ is assumed to obey the equation: 
\begin{equation}
\frac{dS}{dt}=\delta R=\delta (N-S), \label{S}
\end{equation}
because $I$ is almost zero after the infection. The initial value is $S(t_{n\infty})$ in the time evolution of Eq.~(\ref{S}) for  $t_n<t<t_{n+1}$. The linear differential equation Eq.~(\ref{S}) can be solved as $S(t)=N-\{N-S(t_{n\infty})\}e^{-\delta(t-t_n)}$. For $t_n<t<t_{n+1}$, $I(t)$ changes as 
\begin{equation}
I(t)=I(t_n)e^{\int_{t_n}^t (\beta S(t)-\gamma)dt}.  
\end{equation}
$I(t)$ initially decreases since $S(t)$ satisfies $S(t)<\gamma/\beta$ at $t\simeq t_n$. However, $S(t)$ increases with time, and then $I(t)$ increases after the time when $S(t)$ satisfies $\beta S(t)=\gamma$. The $(n+1)$th infection spreading sets in at the time $t=t_{n+1}$ satisfying  
\begin{equation}
\int_{t_n}^t (\beta S(t)-\gamma)dt=\beta\left \{N(t-t_n)-(N-S(t_{n\infty}))\frac{1-e^{-\delta(t-t_n)}}{\delta}\right \}-\gamma (t-t_n)=0.  \label{tm}
\end{equation}
At $t=t_{n+1}=t_{(n+1)0}$, the small but non-negligible infection population $I(t)$ recovers to the initial value $I(t_{n\infty})$, i.e., $I(t_{(n+1)0})=I(t_{n\infty})$. The interval $\Delta t$ between the $n$th and $(n+1)$th infection spreading is numerically evaluated from the relation
\begin{equation}
\Delta t=\{N-S(t_{n\infty})\}\frac{1-\exp(-\delta \Delta t)}{N\delta}-\frac{\gamma}{N\beta}\Delta t.
\end{equation} 
Because that $\beta S(t_{(n+1)0})-\gamma$ takes a positive value at the start time $t=t_{(n+1)0}$ of the $(n+1)$th infection spreading, the infection spreads exponentially again.
 
Thus, the times of the infection spreading $t_n$, the time evolutions of $S(t)$ for $t_n<t<t_{n+1}$, and the peak value of $I$ at the $n$th infection spreading can be successively evaluated. 
\begin{figure}[h]
\begin{center}
\includegraphics[height=4cm]{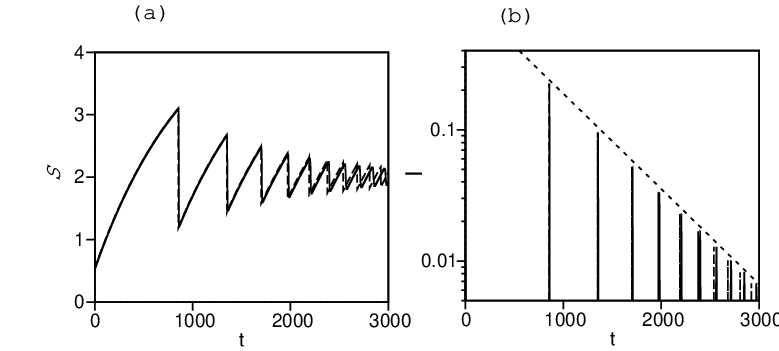}
\end{center}
\caption{(a) Comparison of the direct numerical simulation (solid line) of $S(t)$ with the theoretical approximation (dashed line) at $N=5$, $\beta=2$, $\gamma=4$, and $\delta=0.001$. The initial condition is $S(0)=4.999$ and $I(0)=0.001$. (b) Comparison of the direct numerical simulation (solid line) of $I(t)$ with the theoretical approximation (dashed line) at the same parameter set.}
\end{figure}
Figure 3(a) compares the direct numerical simulation (solid line) of $S(t)$ and the theoretical approximation (dashed line) at $N=5$, $\beta=2$, $\gamma=4$, and $\delta=0.001$. The initial condition is $S(0)=4.999$ and $I(0)=0.001$.    
Figure 3(b) compares the direct numerical simulation (solid line) of $I(t)$ and the theoretical approximation (dashed line) at the same parameter set. 
The dotted line shows the exponential decay $\exp(-0.00167t)$.
Good agreement is seen. However, the timing of the infection spreading deviates  for $t>2500$ partly because the time interval between the start and end of each infection spreading is neglected in the above approximation.

\section{Nonlinear dynamics in the simplified SIRS model with periodic infection rate}
In general, the infection rate can change in time. The SIR model considering the seasonality has been studied by many authors. Complicated dynamics including chaos was found in the SIR model with the time-dependent infection rate~\cite{Aron, Rand}. In this section, we consider the simplified SIRS model with the time-dependent infection rate. We will show that the generalized model can be analyzed with the mapping method. The model equation is written as 
\begin{eqnarray}
\frac{dS}{dt}&=&-(\beta_0+\beta_1\sin\omega t) SI+\delta (N-S-I),\nonumber\\
\frac{dI}{dt}&=&(\beta_0+\beta_1\sin\omega t) SI-\gamma I. \label{MSIR3}
\end{eqnarray}
Here, the infection rate $\beta$ is assumed to change periodically in time as $\beta(t)=\beta_0+\beta_1\sin\omega t$. Figures 4(a), (b), (c), and (d) show the time evolutions of $I(t)$ at $\beta_1=0.1$, $0.2$, $0.21$, and $0.2135$ for $\beta_0=2$, $\gamma=4$, $\omega=2\pi/300$, $\delta=0.001$, and $N=5$. The initial condition is $S(0)=4.999$ and $I(0)=0.001$. Owing to the periodic change of the infection rate, the infection spreading occurs periodically at $\beta_1=0.1$. 
That is, the spike-like infection spreading repeats infinitely, although $I(t)$ approaches a constant value at $t=\infty$ in case of $\beta_1=0$. As $\beta_1$ increases, more complicated nonlinear phenomena such as the period-doubling bifurcation and chaos appear. A period-2 solution appears in Fig.~4(b), and a period-4  solution appears in Fig.~4(c). The chaotic time evolution  appears in Fig.~4(d). 
We have calculated the Lyapunov exponent of Eq.~(\ref{MSIR3}) for $\beta_0=2$, $\gamma=4$, $\omega=2\pi/300$, $\delta=0.001$, and $N=5$.
Figure 5 shows the Lyapunov exponent of Eq.~(\ref{MSIR3}) as a function of $\beta_1$. The period-doubling bifurcation occurs at $\beta_1\simeq 0.2054$ and the period-2 solution appears. The Lyapunov exponent takes 0 at the critical parameter of the period-doubling bifurcation. The period-doubling bifurcation from the period-2 solution to the period-4 solution occurs at $\beta_1\simeq 0.2099$, and the period-doubling bifurcation from the period-4 solution to the period-8 solution occurs at $\beta_1\simeq 0.21085$.   The chaos appears for $\beta_1>0.2111$ where the Lyapunov exponent takes a positive value. 
\begin{figure}[h]
\begin{center}
\includegraphics[height=3.5cm]{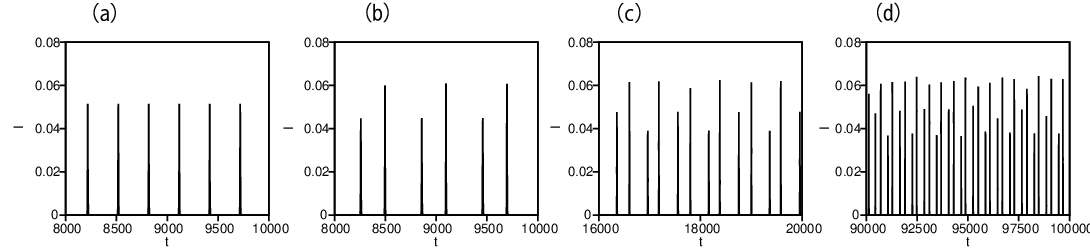}
\end{center}
\caption{Time evolutions of $I(t)$ at (a) $\beta_1=0.1$, (b) $0.2$, (c) $0.21$, and (d) $0.2135$ for $\beta_0=2$, $\gamma=4$, $\omega=2\pi/300$, $\delta=0.001$, and $N=5$ in Eq.~(\ref{MSIR3}).}
\end{figure}
\begin{figure}[h]
\begin{center}
\includegraphics[height=4cm]{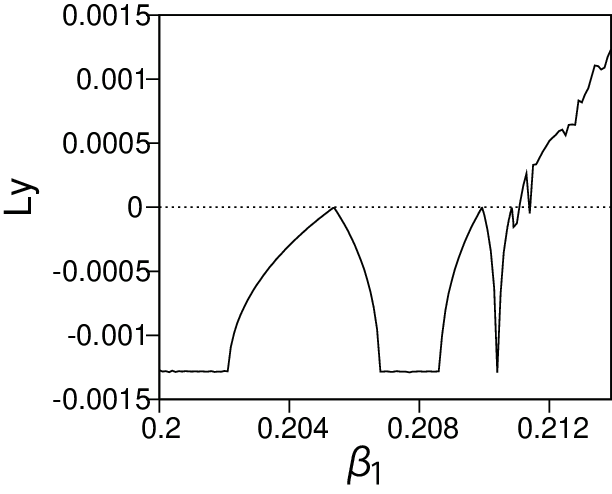}
\end{center}
\caption{Lyapunov exponent of Eq.~(\ref{MSIR3}) as a function of $\beta_1$ for $\beta_0=2$, $\gamma=4$, $\omega=2\pi/300$, $\delta=0.001$, and $N=5$.}
\end{figure}

When $\delta$ is sufficiently small, the approximation in the previous section can be applied. $S(t_{n\infty})$ at the final time of the $n$th infection spreading is determined by the initial value $S(t_{n0})$ of the $n$th infection spreading from the equation: 
\begin{equation}
S(t_{n\infty})-(\gamma_0+\gamma_1\sin\omega t_n)/\beta\log S(t_{n\infty})=S(t_{n0})-(\gamma_0+\gamma_1\sin\omega t_n)/\beta\log S(t_{n0}) \label{SF2}.
\end{equation}
$S(t)$ increases as $S(t)=N-\{N-S(t_{n\infty})\}e^{-\delta(t-t_n)}$ for $t_n<t<t_{n+1}$, and $I(t)$ changes as
\begin{equation}
I(t)=I(t_n)e^{f(t)},
\end{equation}
where 
\[f(t)=\int_{t_n}^t \{(\beta_0+\beta_1\sin\omega t)S(t)-\gamma\}dt.\]
$f(t)$ can be explicitly expressed as
\begin{eqnarray}
f(t)&=&\beta_0N(t-t_n)+\beta_1N\omega(-\cos\omega t+\cos\omega t_n)-\beta_0\{N-S(t_{n\infty})\}\frac{1-e^{-\delta(t-t_n)}}{\delta}\nonumber\\
&+&\frac{e^{-\delta(t-t_n)}}{\delta^2+\omega^2}(-\delta\sin\omega t-\omega\cos\omega t)-\frac{1}{\delta^2+\omega^2}(-\delta\sin \omega t_n-\omega\cos\omega t_n)-\gamma(t-t_n)\nonumber\\ 
\end{eqnarray}
The interval $\Delta t$ between the $n$th and $(n+1)$th infection spreading is numerically evaluated from $f(t_{n+1})=0$. The initial value  $S(t_{(n+1)0})$ at the $(n+1)th$ infection spreading is given by $N-\{N-S(t_{n\infty})\}e^{-\delta(t_{n+1}-t_n)}$, and the peak value of $I$ at the $(n+1)$th infection spreading is evaluated at 
\[S(t_{(n+1)0})-(\gamma/\beta) \log S(t_{(n+1)0})-\gamma/\beta+(\gamma/\beta)\log(\gamma/\beta),\]
where $\beta=\beta_0+\beta_1\sin\omega t_{n+1}$.
The map from $S(t_{n0})$ to $S(t_{(n+1)0})$ is determined by these relations. 
Figures 6(a), (b), (c), and (d) show the time evolutions of $I(t_n)$ at $\beta_1=0.1$, 0.2, 0.21, and 0.1235 obtained by the map. The other parameters are set to be $\beta_0=2$, $\gamma=4$, $\omega=2\pi/300$, $\delta=0.001$, and $N=5$.
The period-doubling bifurcation and chaos are reproduced in this map. 
\begin{figure}[h]
\begin{center}
\includegraphics[height=3.5cm]{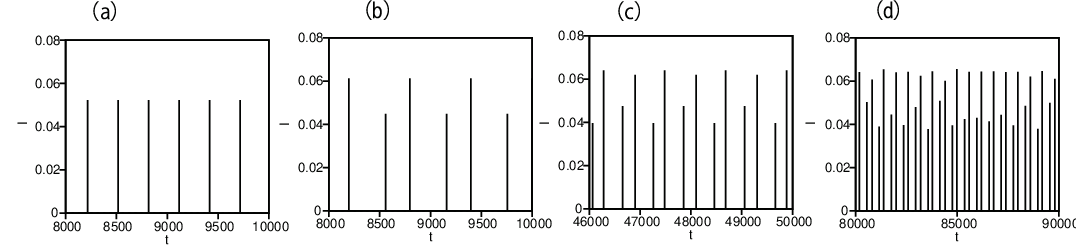}
\end{center}
\caption{Time evolutions of $I(t)$ at (a) $\beta_1=0.1$, (b) $0.2$, (c) $0.21$, and (d) $0.2135$ for $\beta_0=2$, $\gamma=4$, $\omega=2\pi/300$, $\delta=0.001$, and $N=5$ by the approximate map.}
\end{figure}
\section{Directed evolution in the coupled SIRS model with mutation on a one-dimensional lattice} 
In the COVID-19 pandemic, new types of viruses such as $\beta$, $\delta$, and $o$ appeared successively. It is known that infectivity tends to increase, and virulence tends to decrease in the process of mutation and evolution. The evolution of infectivity and virulence has been studied by many authors~\cite{Anderson, May2, Sasaki}. 
We propose a simple model equation to study the mutation and evolution process using the simplified SIRS model. It has a form of coupled equations of the simplified SIRS models. The population infected by the $i$th type of virus is expressed as $I_i(t)$ and the corresponding recovered population is denoted by $R_i(t)$. The susceptible population has no type and is expressed as $S(t)$. The mutation occurs in the genetic sequence, that is, the base substitution in DNA is a fundamental process of the mutation. Although the diffusion in the hyper-space of the base sequence is more realistic, we consider a one-dimensional lattice model for the sake of simplicity. That is, we assume that the $(i+1)$th and $(i-1)$th types appear from the $i$th type with mutation rate $D$. In other words, the population $I_i$ increases by $D(I_{i+1}+I_{i-1})$ owing to the mutation from the $(i+1)$th and $(i-1)$th types and decreases by $2DI_i$ due to the mutation of the $i$th type. The infection rate $\beta$ depends on the type and is expressed as $\beta_i$. As a simple model, $\beta_i$ is assumed to be $\beta_i=\beta_0+\beta_1(i-L/2)$, where $L$ is the total number of types. That is, the infection rate is assumed to be higher for larger $i$. The initial type is $L/2$ and the infection rate is $\beta_0$. Thus, the model equation is written as 
\begin{eqnarray}
\frac{dS}{dt}&=&-\sum \beta_i SI_i+\delta \sum R_i,\nonumber\\
\frac{dI_i}{dt}&=&\beta_i SI_i-\gamma I_i+D(I_{i+1}-2I_{i}+I_{i-1}),\nonumber\\
\frac{dR_i}{dt}&=&\gamma I_i-\delta R_i. \label{DSIR}
\end{eqnarray}
The total population $N=S+\sum I_i+\sum R_i$ is conserved. 
The boundary conditions are $I_0=I_1$ and $I_{L+1}=I_{L}$.
Equation ~(\ref{DSIR}) takes the form of coupled equations of the simplified SIRS models which can also be interpreted as a discrete-type reaction-diffusion equation. 

\begin{figure}[h]
\begin{center}
\includegraphics[height=4cm]{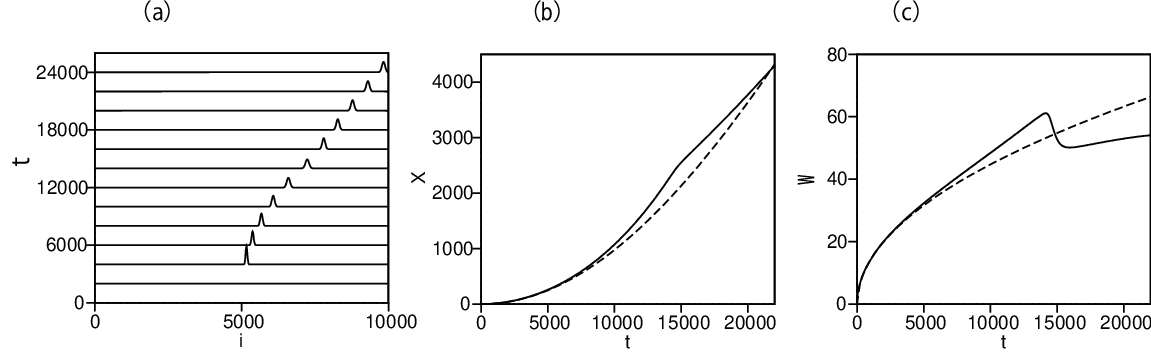}
\end{center}
\caption{(a) Time evolution of the profile $I_i$ by Eq.~(\ref{DSIR}) for $\beta_0=2$,$\beta_1=0.00005$, $\gamma=4$, $\delta=0.001$, $D=0.1$, and $L=10000$. The initial condition is $S=4.999$ and $I_{L/2}=0.001$. (b) Time evolution of the average displacement $X$. (c) Time evolution of the width $W$. }
\end{figure}

Figure 7(a) is the time evolution of the profile $I_i$ for $\beta_0=2$, $\beta_1=0.00005$, $\gamma=4$, $\delta=0.001$, $D=0.1$, and $L=10000$. The initial condition is $S=4.999$, $I_{L/2}=0.001$, $I_i=0$ for $i\ne L/2$, and $R_i=0$. 
A pulse-like wave propagates in the right direction. To characterize the pulse-like wave, we calculated the average position $\langle i(t)\rangle$ and the width $W(t)$: 
\[\langle i(t)\rangle =\sum i I_i/\sum I_i,\;\; W(t)=\{(\sum i^2I_i/\sum I_i)^2-\langle i(t)\rangle^2\}^{1/2}.\]
Figures 7(b) and 7(c) shows the time evolution of the average displacement $X(t)=\langle i(t)\rangle-\langle i(0)\rangle$ and width $W(t)$ of the pulse-like wave. The dashed line in Fig.~7(c) is $\sqrt{2Dt}$. That is, the width $W$ grows as $t^{1/2}$ initially, $W$ decreases around $t=15000$, and tends to saturate for $t>15000$. The displacement $X$ increases as $t^2$ initially, but $X$ seems to change in proportion to $t$ for $t>15000$. 
For sufficiently large $t$, $X$ takes a constant value close to $L/2$, that is, the evolution stops at the boundary. This simple model reproduces the evolution toward the type of higher infection rate. 
We do not understand the reason of the non-smooth dynamical behavior of $W$ as shown in Fig.~7(b) yet. The rapid increase of $X$ might cause the non-smooth behavior. 

In the stationary state of Eq.~(\ref{MSIR}), $S$ takes a value $\gamma/\beta$.
Under this approximation, $I_i$ obeys an equation: 
\[\frac{dI_i}{dt}=(\beta_i \gamma/\beta-\gamma)I_i+D(I_{i+1}-2I_{i}+I_{i-1}).\]
However, the total sum $\sum I_i$ is not conserved in this model equation. A simple model satisfying the conservation law is
\begin{equation}
\frac{dI_i}{dt}=(\beta_i\gamma/\beta-\langle \beta_i\gamma/\beta\rangle)I_i+D(I_{i+1}-2I_{i}+I_{i-1}),  \label{ev}
\end{equation}
where $\langle \beta_i\gamma/\beta\rangle=\sum (\beta_i\gamma/\beta)I_i/\sum I_i$ is the weighted average of the effective infection rate $\beta_i\gamma/\beta$ by the weight $I_i$. The constant part of the effective infection rate independent of $i$ such as $-\gamma$ does not appear in this model equation because only the difference of the effective infection rate from the weighted average determines the dynamics. The total sum of $I_i$ is conserved in the time evolution by Eq.~(\ref{ev}). The first term of Eq.~(\ref{ev}) implies that the population $I_i$ with the effective infection rate larger than the weighted average grows in time. This type of evolution equation was proposed by one of the authors~\cite{Sakaguchi2}.
A similar pulse-like wave appears even in Eq.~(\ref{ev}).  
\begin{figure}[h]
\begin{center}
\includegraphics[height=4cm]{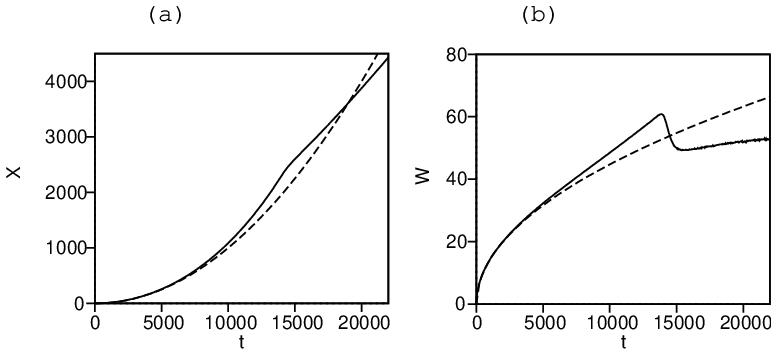}
\end{center}
\caption{Time evolutions of (a) the average displacement $X$ and (b) width $W$ by Eq.~(\ref{ev}) for $\gamma/\beta=2$, $\beta_i=2+0.00005(i-L/2)$, $D=0.1$, and $L=10000$.  }
\end{figure}
Figures 8(a) and (b) are the time evolutions of the average displacement $X$ and width $W$ for $\gamma/\beta=2$, $\beta_i=2+0.00005(i-L/2)$, $D=0.1$, and $L=10000$ for Eq.~(\ref{ev}). 
The initial condition is $I_{L/10}=0.001$ and $I_i=0$ for $i\ne L/10$. 
Time evolutions of $X$ and $W$ shown in Figs.~8(a) and (b) are similar to those shown in Figs.~7(b) and (c). The dashed line in Fig.~8(b) is $\sqrt{2Dt}$. The time evolution of $W$ begins to deviate from the square root law near $t=5000$. The average displacement initially grows as $t^2$ but changes in proportion to $t$ for $t>15000$. 

The continuum approximation of Eq.~(\ref{ev}) yields
\begin{equation} 
\frac{\partial I}{\partial t}=\{\beta(x)\gamma/\beta-\langle \beta(x)\gamma/\beta\rangle\}I+D\frac{\partial^2I}{\partial x^2}.  \label{ev2}
\end{equation}
Here, $\beta(x)=\beta_0+\beta_1(x-L/2)$. There is a Gaussian-type solution to Eq.~(\ref{ev2}): $I(x)=A\sqrt{\alpha(t)/\pi}\exp[-\alpha(t)\{x-X(t)\}^2]$, where $\alpha(t)$ and $X(t)$ satisfy
\begin{equation}
\frac{d\alpha}{dt}=-4D\alpha^2,\;\;\; \frac{dX}{dt}=\frac{\gamma\beta_1}{2\beta\alpha}. \label{ev3}
\end{equation}
The solution of Eq.~(\ref{ev3}) is 
\[\alpha(t)=\frac{\alpha(0)}{1+4\alpha(0)Dt}, X(t)=X(0)+\frac{\gamma\beta_1}{2\beta\alpha(0)}t+(\gamma/\beta)\beta_1Dt^2,\]
where $\alpha(0)$ and $X(0)$  are initial values of $\alpha(t)$ and $X(t)$. The width $W$ of the Gaussian function is expressed with $\alpha$ as $W=1/(2\alpha(t))$. If the initial condition is $I(x)=\delta(x)$, $W=\sqrt{2Dt}$ and $X=(\gamma/\beta)\beta_1Dt^2$. The dashed line in Fig.~8(a) is $(\gamma/\beta)\beta_1Dt^2$ and the dashed line in Fig.~8(b) is $W=\sqrt{2Dt}$. The dashed line in Fig.~7(b) is $X=\gamma/(\beta_0+\beta_1X_0)\beta_1Dt^2$ where $X_0$ is the first approximation of $X$: $X_0=(\gamma/\beta_0)\beta_1Dt^2$ and $\beta$ is approximated at $\beta_0+\beta_1 X_0$. 
The continuum approximation is initially good, but the deviation is observed for $t>5000$. Similar behavior was observed in Eq.~(\ref{DSIR}) as shown in Figs.~7(b) and (c). 

\section{Summary}   
We studied the nonlinear dynamics of two simple epidemic models based on the SIRS model. In Sec.~2 and 3, we studied the repetitive spreading of infection in a simplified SIRS model. If $\delta$ is zero, the model equation is reduced to the SIR model, and the infection spreading occurs only once. Even if $\delta$ is sufficiently small, the infection spreading occurs many times, that is, the dynamical behavior changes drastically owing to the terms $\delta R$. We showed that the time evolution can be reduced to a map when the decay rate $\delta$ of acquired immune is sufficiently small. The repetitive spreading of infection was reproduced by the map. The approximation can also be applied to the SIRS model with the periodic variation of infection rate. The map reproduced the period-doubling bifurcation and chaos. In Sec.~4, we studied a simple evolution model which has a form of the coupled SIRS equations where the mutation is represented by a discrete-type diffusion equation. We confirmed that the strengthening of infection rate and weakening of virulence occur in this type of simple model of evolution. By the continuum approximation, a kind of reaction-diffusion equation for $I$ is derived. The reaction-diffusion equation with the infection rate depending linearly on $x$ has an exact solution. The initial time evolutions in the discrete models Eqs.~(\ref{DSIR}) and (\ref{ev}) were well approximated at the solution of Eq.~(\ref{ev2}). 

The main results in this paper are that the nonlinear dynamics and the evolution can be reproduced respectively by the map derived from the differential equation and the exact solution of the reaction-diffusion equation. We studied the SIRS model and the coupled SIRS model for several parameter sets. We would like to study the model equations further in the future.

\end{document}